\documentclass[10pt]{iopart}
\usepackage{bbm,mathrsfs}
\usepackage{graphicx}
\usepackage{amsfonts}
\usepackage{iopams}

\def\textbf#1{{\bf #1}}
\def\be{\begin{equation}}
\def\ee{\end{equation}}
\def\ben{\begin{eqnarray}}
\def\een{\end{eqnarray}}
\def\eea{\end{array}}
\def\bea{\begin{array}}
\newcommand{\ot}[0]{\otimes}
\newcommand{\bei}{\begin{itemize}}
\newcommand{\eei}{\end{itemize}}
\newcommand{\ket}[1]{|#1\rangle}
\newcommand{\bra}[1]{\langle#1|}

\begin{document}

\title[Quantum kinetic Ising models]{Quantum kinetic Ising models}

\author{R Augusiak}

\address{ICFO--Institut de Ci\`encies Fot\`oniques, Parc Mediterani de la Tecnologia,
 08860 Castelldefels (Barcelona), Spain}
\ead{remigiusz.augusiak@icfo.es}

\author{F M Cucchietti}
\address{ICFO--Institut de Ci\`encies Fot\`oniques, Parc Mediterani de la Tecnologia,
 08860 Castelldefels (Barcelona), Spain}
 \ead{fernando.cucchietti@icfo.es}

\author{ F Haake}
\address{Fachbereich Physics, Universit\"at Duisburg--Essen, 47048 Duisburg, Germany}
\ead{fritz.haake@uni-duisburg-essen.de}

\author{M Lewenstein}
\address{ICFO--Institut de Ci\`encies Fot\`oniques, Parc Mediterani de la Tecnologia,
 08860 Castelldefels (Barcelona), Spain}
\address{ICREA--Instituci\'o Catalana  de Recerca i Estudis Avan\c cats,
Lluis Companys 23, 08010 Barcelona, Spain}
\ead{maciej.lewenstein@icfo.es}

\begin{abstract}
We introduce a quantum generalization of classical kinetic Ising
models, described by a certain class of quantum many body master
equations. Similarly to kinetic Ising models with detailed balance
that are equivalent to certain Hamiltonian systems, our
models reduce to a set of Hamiltonian systems determining  the
dynamics of the elements of the many body density matrix. The
ground states of these Hamiltonians are well described by
matrix product, or pair entangled projected states. We discuss
critical properties of such Hamiltonians, as well as entanglement
properties of their low energy states.
\end{abstract}

\maketitle

\section{Introduction}
The paper that we have prepared for the special issue of the New
Journal of Physics on "Quantum Information and Many-Body Theory"
has two apparently independent motivations. First, it is motivated
by the recent interest in many body quantum master equations and
design of open systems for quantum state engineering and
quantum simulations. Second, it concerns entanglement properties
of quantum many body states corresponding to classical kinetic
models with detailed balance and their generalizations to
quantum master equations.

\subsection{Quantum master equations}

The quantum master equation (QME) is the basic theoretical tool to
describe the evolution of open systems that undergo Markovian
dynamics \cite{fritz,quantumnoise}. While the initial studies of
QMEs dealt with many body applications, over the last decades most
research has focused on systems of few degrees of freedom, mainly
because of the unquestionable complexity of the many body QMEs. In
the recent fifteen years, however, there have been two waves of
renewed interest in many body QMEs, related to the unprecedented
progress of experimental control and engineering of ultracold
atomic and molecular systems. On one hand, there has been a wealth
of interest in the studies of various kind of cooling processes
using QMEs. Zoller, Gardiner, and collaborators
\cite{zoller-gardiner} studied in a series of papers the growth of
a Bose-Einstein condensate in a trapped bosonic gas. One of us
(M.L.), together with I. Cirac, P. Zoller, Y. Castin,  and others,
developed the theory of laser cooling of Bose gases
\cite{lasercool}, in particular in the so called {\it festina
lente} limit \cite{prl-castin}. Similar ideas were applied to the
processes of laser cooling of Fermi gases \cite{lasercool-fermi}
and sympathetic cooling \cite{clz,nova0,nova}.

On the other hand, several authors proposed to make use of the
capabilities of modern quantum optics and atomic physics
experimental methods to {\it design} and {\it realize
experimentally} QMEs that allow to engineer interesting quantum
many body {\it pure} states
\cite{cirac-qme,zoller-qme1,zoller-qme2}. These pure states range
from simple Bose Einstein condensates to the stabilizer states,
MPS-PEPS states, or states with topological order. The first
experiment realizing these ideas was conducted in the group of G.
Rempe \cite{rempe}, who has been able to prepare a 1D bosonic
Tonks--Girardeau \cite{tonks-girardeau} gas employing three-body
losses. So far most of these new proposals concern ultracold
atoms, molecules, in particular polar molecules,  or ions, but
recently they start to involve Rydberg atoms, that are
particularly suitable for design and realization of 3-body, 4-body
etc. interactions thanks to the Rydberg blockade mechanism. Weimer
{\it et al.} \cite{weimer}, proposed to use  many-body quantum
gates stroboscopically, employing the Rydberg blockade effect to
engineer the topologically ordered ground state of the famous
Kitaev toric code, the color code \cite{kitaev}, and even to
realize a quantum simulator of the $U(1)$ lattice gauge theory.

Motivated by these developments we propose in this paper a new
class of many body QMEs generalizing classical kinetic models (in
particular kinetic Ising models (KIMs); for a review see
\cite{kawasaki}). Our QMEs have as stationary states thermal
Boltzmann--Gibbs states of the underlying classical model. Note
that these thermal states might correspond to very complex quantum
states, for instance if the underlying classical model concerns
kinetics of commuting stabilizer operators for the cluster states
\cite{briegel}, as we discuss in Sec. \ref{conclusions}. The
diagonal elements of the density matrix in our models undergo
dynamics equivalent to that of the underlying classical model. The
off-diagonal elements of the density matrix exhibit complex
evolution, but fortunately can be grouped into independently
evolving blocks. Similarly to kinetic Ising models with detailed
balance, which are equivalent to certain Hamiltonian systems, our
models reduce to a set of Hamiltonian systems determining  the
dynamics of the elements of the many body density matrix. In 1D we
identify classes of these models that are exactly soluble. In
general, the ground states of these Hamiltonians are well
described by the matrix product states (MPS), or pair entangled
projected states (PEPS) \cite{david,wolf}.

It should be noticed that there exist previous efforts in the
above direction. For instance, in Ref. \cite{Heims} (see also Ref.
\cite{kawasaki}) a general master equation for kinetic Ising model
with an environment was derived. Under some assumptions, the
diagonal elements of these formulations also reproduce Glauber's
kinetic model, although no full solution of the equation was
attempted.

\subsection{Entanglement in many body systems}

The studies of the role of entanglement in many body systems were
definitely initiated by the seminal Ref. \cite{fazio}, but they go
back to the early works on area laws in quantum systems
\cite{srednicki} (for a review see \cite{fazio-rmp,wolf,eisert}.
Let us remind the readers that grounds states of (non-critical)
many body systems with local Hamiltonians exhibit the area law.
This means that if we divide the whole system into a subsystem $A$
and the rest $B$ where the size $|A|$ of $A$  is large, but much
smaller that that of $|B|$, and calculate the von Neumann entropy
of the reduced density matrix of $A$, the latter will scale as the
size of the boundary ("area") of $A$, $S_A\varpropto |\partial
A|$. The area law expresses the fact that away from criticality,
correlations --and in particular quantum correlations responsible
for entanglement-- decay on short length scales.

The situation is different at criticality, although so far only 1D
systems are fully understood. In one dimension, the area law at
criticality may get logarithmic corrections so that the entropy of
the block of size $L$ scales as $S_L\varpropto c\log(L)$, where
the constant $c$ can be related to the charge of the conformal
field theory describing the corresponding critical behavior. In
higher dimensions the situation is much less clear, and no
universal laws at criticality are known \cite{wolf}. Very
recently, Masanes \cite{masanes} proved under quite general
conditions that for the ground states of systems with local
Hamiltonians, the entropy of a block $A$ is bounded from above by
$|A|\log{A}$.

There is a class of quantum states that always fulfill the area
law in any dimension, despite the fact that they exhibit
criticality for certain values of parameters \cite{wolf}. These
states are related to thermal states of classical Hamiltonians,
such as Ising, or Potts models. In fact, any set of local quantum
mechanical commuting operators can be used to build such models.
For example, one can take stabilizer operators for cluster states
\cite{briegel}, or star and plaquette operators of the Kitaev
model \cite{kitaev}. In the following we will mainly focus on
Ising models, leaving the discussion of more complicated cases to
further publications.

For Ising models, the states in question have the form\footnote{We
would like to warn our readers against a possible
misunderstanding: At issue are irreversible processes whose
quantum behavior in general involves mixed states described by
density operators. The generators in the pertinent master
equations are non-Hermitian operators, precisely due to the
irreversible dynamics. However, given detailed balance a master
equation can be rewritten such that the generator becomes
Hermitian and is then called a "Hamiltonian". Following widespread
practice we shall  indulge in calling a so transformed master
equation a "Schr\"odinger equation in imaginary time" and talking
about eigenstates of the "Hamiltonian"; the resulting formalism
then even allows to deal with superposition states, Eq.
(\ref{states}) being the first example. Nevertheless, this paper
is about density operators and damped motion. The "state" in Eq.
(\ref{states}) represents the canonical density operator.}

\begin{equation}\label{ground}
|\Psi\rangle=\frac{1}{\sqrt{Z_{N}}}\sum_\sigma \exp[-\beta
H(\sigma)/2]|\sigma\rangle, \label{states}
\end{equation}
where $\sigma=(\sigma_1, \ldots,\sigma_N)$ denotes a configuration
of $N$ Ising spins ($\ket{\sigma}$ stands for a vector
representing $\sigma$ in the corresponding  Hilbert space),
$H(\sigma)$ is the corresponding classical Hamiltonian, and
$Z_{N}=\Tr\exp[-\beta H(\sigma)]$ is the partition function. As
pointed out in Refs. \cite{david,wolf}, and references therein,
these states have the following properties:
\begin{itemize}
\item They are associated to a family of classical kinetic
Ising  models (KIMs) that describe the approach to the classical
thermal equilibrium state $\exp[-\beta H(\sigma)]/Z_{N}$ and obey
detailed balance.

\item The KIMs in question can be transformed by an appropriate ansatz
into an equivalent problem of Hamiltonian dynamics in imaginary
time (i.e. describing decay in time), with a certain quantum
Hamiltonian $\hat H$ parametrically depending on $\beta$.

\item The Hamiltonian $\hat H$ is non-negative and has one
eigenvector given by the expression (\ref{states}) corresponding to the
eigenvalue zero. Away from criticality $\hat H$ is gapped, i.e.
excited states have strictly positive energies. As $\beta \to
\beta_c$ the gap vanishes, and at $\beta_c$ the Hamiltonian is
gapless. The way the gap vanishes determines the dynamical
critical exponent $z$ of the associated KIM.

\item The ground state (\ref{states}) fulfills the area law and does
not exhibit any special behavior at criticality. It can be exactly
represented as a MPS in 1D, or as a PEPS in higher dimensions
\cite{david}.
\end{itemize}

Intrigued by the fact that the states (\ref{states}) always
fulfill the area law, we have attempted to look more closely into
their properties, and the properties of their parent Hamiltonians
$\hat H$. One possible way is to study the entanglement properties
of the excited states of $\hat H$, but we have chosen another
approach. We have generalized the KIMs to quantum models by
defining a new class of QMEs, as explained in the previous
subsection. These QMEs define new classes of parent Hamiltonians,
and the grounds states of these Hamiltonians are expected to be
again well described by matrix product or pair entangled projected
states. This paper is devoted to the discussion of the  critical
properties of such Hamiltonians, as well as entanglement
properties of their low energy states.

\subsection{Plan of the paper}

The paper is organized as follows. In Secion 2 we remind the
readers of basics of KIMs with detailed balance, associated
Hamiltonians, and more. In Section 3 we present the main result of
this paper: we describe the generalization of KIMs to QMEs, and we
discuss the properties of the QMEs. In particular, we show how the
equations for $2^N\times 2^N$ matrix elements of the density
matrix split into $2^N$ equations of the KIM-type for functions of
$2^N$ spin configurations. Detailed balance allows to transform
these quasi-KIM equations to a Hamiltonian form, and we derive
here the new classes of Hamiltonians. Properties of these
Hamiltonians and  their low energy states are discussed in Section
4. Here we show that Hamiltonians describing the evolution of the
off-diagonal density matrix elements (coherences) are strictly
positive\footnote{An operator $O\in\mathscr{B}(\mathscr{H})$ for
some Hilbert space $\mathscr{H}$ is said to be positive if
$\langle\psi|O|\psi\rangle \geq 0$ for any
$\ket{\psi}\in\mathscr{H}$. Then we say that $O$ is strictly
positive if $\langle\psi|O|\psi\rangle > 0$ for any
$\ket{\psi}\in\mathscr{H}$. For a Hermitian operator $O$ acting on
finite dimensional Hilbert space, both conditions are equivalent
to nonnegativity and strict positivity of eigenvalues of $O$.},
which implies that the corresponding coherences decay to zero. At
criticality we observe effects of critical slowing down of
coherences; the Hamiltonians become gapless, implying that at
criticality the decay is of the form of an exponential decay times
algebraic tails. In some cases, criticality may even lead to
survival of coherences for infinite times.  In Section 4 we also
discuss entanglement properties of the ground and excited states
of the corresponding Hamiltonians. Our conclusions and outlook are
contained in Section 5.

\section{Basics of kinetic Ising models}

As we discuss in Section 4, the methods developed in this paper
can be straightforwardly generalized and applied to models other
than Ising models: Potts models, classical clock models, or models
employing commuting stabilizer operators. For simplicity and
concreteness we will limit the discussion to Ising models, with
Ising variables $\sigma_i=\pm 1$, described quantum mechanically
by the commuting Pauli matrices $\sigma^z_i$. We will consider
systems with a classical Hamiltonian $H(\sigma)$ following
Markovian dynamics toward the thermal equilibrium state. The
dynamics for a Markovian stochastic process is most conveniently
formulated for the conditional probability
$P(\sigma,t|\sigma_0,0)$ for the configuration $\sigma$ at time
$t$, provided the initial configuration was $\sigma_0$ at $t=0$.
This conditional probability allows to calculate all multi-time
correlation functions of the process. Note that
$P(\sigma,0|\sigma_0,0) = \delta_{\sigma\sigma_0}$, i.e. at the
initial time conditional probability is obviously given by the
Kronecker delta of the configurations $\sigma$ and $\sigma_0$. In
the following, in order to avoid too many arguments we will
consider dynamics of probability of configurations $P(\sigma,t)$,
which fulfills the same equation as $P(\sigma,t|\sigma_0,0)$, but
with a more general initial condition.

\subsection{Kinetic Ising models with detailed balance}

Here, we define in a more detailed way the classical kinetic Ising
models and recall some of the literature results concerning
particular examples of such models. In general, kinetic Ising
models are defined by specifying the so called kinetic (master)
equation for probability $P(\sigma,t)$ of the form
\begin{equation}\label{classical_master}
\dot P(\sigma,t)=\sum_{\sigma'}\left[w(\sigma'\rightarrow\sigma)P(\sigma',t)
-w(\sigma\rightarrow\sigma')P(\sigma,t)\right],
\end{equation}
where the sum runs over all possible configurations $\sigma$. The
function $w(\sigma\rightarrow\sigma')$, hereafter also called
transition probability, stands for the probability per unit time
that the system changes its configuration from $\sigma$ to
$\sigma'$.

Usually one imposes the so called detailed balance condition (DBC)
that says that at equilibrium the probability per unit time of a
transition from $\sigma$ to $\sigma'$ is the same as the
probability of transition in the opposite direction,
$\sigma'\to\sigma$. Mathematically, it reads
\begin{equation}\label{detailed_balance}
w(\sigma'\rightarrow\sigma)P_{\mathrm{eq}}(\sigma')=
w(\sigma\rightarrow\sigma')P_{\mathrm{eq}}(\sigma),
\end{equation}
where $P_{\mathrm{eq}}(\sigma)$ is the equilibrium probability
distribution $P_{\mathrm{eq}}(\sigma)=P(\sigma,t\to\infty)$.

To support the above general remarks with some more detailed
investigations let us now discuss some previously studied examples
of such kinetic models. At the beginning let us focus on the
one-dimensional Ising spin system, i.e., a chain of $N$ spins
uniformly distributed on a line. In this case $\sigma$ denotes one
of the $2^{N}$ possible configurations of $N$ spins and can be
represented as a $N$--dimensional vector
$(\sigma_{1},\ldots,\sigma_{N})$ with $\sigma_{i}=\pm1$
$(i=1,\ldots,N)$.

We restrict our attention to the case in which the behavior of
$i$th spin is local, i.e., depends only on the nearest neighbors
(generalizations to local models with next nearest neighbors are
straightforward). We also assume  that the probability
distribution at equilibrium is
\begin{equation}\label{IsingEquilibrium}
P_{\mathrm{eq}}(\sigma)=\frac{1}{Z_{N}}\mathrm{e}^{-\beta\mathcal{H}(\sigma)}
\end{equation}
with $\mathcal{H}$ denoting the classical (local) Ising
Hamiltonian. In particular, we consider here the ferromagnetic
Ising model which in 1D corresponds to
\begin{equation}\label{IsingHamil}
\mathcal{H}(\sigma)=-J\sum_{i}\sigma_{i}\sigma_{i+1}\qquad (J>0).
\end{equation}
In this case the partition function has the explicit form
$Z_{N}=2^{N}(\cosh^{N}\beta J+\sinh^{N}\beta J)$. The simplest
possible process that may occur here is a single flip of the $i$th
spin. Schematically this process can be stated as $\sigma\to
D_{i}\sigma$, where $D_{i}$ denotes the flip at $i$th position,
$D_{i}\sigma=(\sigma_{1},\ldots,-\sigma_{i},\ldots,\sigma_{N})$.
Also, let $w(\sigma\to D_{i}\sigma)$ denote the transition
probability for that process.
The only processes that lead to the configuration $\sigma$,
appearing on the left--hand side of Eq. (\ref{classical_master})
are spin flips $D_{i}\sigma\to \sigma$ for any $i=1,\ldots,N$. The
inverse type of processes can drag the system away from $\sigma$.
This means that the general kinetic equation
(\ref{classical_master}) can be reduced in this case to a much
simpler form
\begin{eqnarray}\label{1DGlauber2}
\dot{P}(\sigma,t)&=&\sum_{i=1}^{N}\left[w(D_{i}\sigma\to\sigma)
P(D_{i}\sigma,t)-w(\sigma\to D_{i}\sigma)P(\sigma,t)\right].
\end{eqnarray}
The most general form of $w(\sigma\to D_{i}\sigma)$ in the case
that the interaction with both nearest neighbors is symmetric and
leads the system to the equilibrium state (\ref{IsingEquilibrium})
was shown to be \cite{Glauber}:
\begin{equation}\label{spinrates}
w(\sigma\to D_{i}\sigma)=\Gamma(1+\delta\sigma_{i-1}\sigma_{i+1})
\left[1-\textstyle\frac{1}{2}\gamma\sigma_{i}(\sigma_{i-1}+\sigma_{i+1})\right],
\end{equation}
where $\gamma=\tanh2\beta J$ (notice that the value $\gamma=0$
corresponds to infinite temperature, while $\gamma=1$ to zero
temperature), $|\delta|\leq 1$ and $0<\Gamma<\infty$.

The case of $\delta=0$ was thoroughly investigated by Glauber, and
shown to be solvable in the sense that all the relevant physical
quantities can be computed analytically. In particular, the
non--equilibrium expectation values and equilibrium correlation
functions were determined. Moreover, this model was shown to have
the dynamical exponent\footnote{This means that the time dependent
spin--spin correlation function decay on time scale as
$t_{\mathrm{dec}}$ behaves as $t_{\mathrm{dec}}\propto\xi^{z}$,
where $z$ is the dynamical critical exponent, and $\xi$ denotes
the correlation length; $\xi$ scales as $(1-\gamma)^{-1/2}$ when
$\gamma\to 1$.} $z=2$. Later on, the above model was treated in a
series of papers \cite{DekerHaake,Kimball,HaakeThol} for nonzero
$\delta$. In particular, it was shown in Refs.
\cite{DekerHaake,Kimball} that the choice
$\delta=\gamma/(2-\gamma)$ entails the dynamical exponent $z\neq
2$.

KIMs have also been studied in two and higher dimensions, although
in these cases there is no known analytical solution. However,
using efficient techniques precise numerical results have been
obtained for relevant quantities such as the critical dynamical
exponent \cite{TwoDKIM}.

\subsection{Associated quantum Hamiltonians}
\label{Acciociated}

Interestingly, as it was show in e.g. Refs.
\cite{Glauber,Felderhof,Felderhof2} the detailed balance condition
(\ref{detailed_balance}) allows to rewrite the master equation
(\ref{classical_master}) as a Schr\"odinger equation. In order to
see that let us introduce the function $\phi(\sigma,t)$ related to
the probability $P(\sigma,t)$ through
$P(\sigma,t)=\sqrt{P_{\mathrm{eq}}(\sigma)}\,\phi(\sigma,t)$,
where as above $P_{\mathrm{eq}}$ is the equilibrium distribution.
Inserting the latter in the master equation
(\ref{classical_master}) and reorganizing slightly some terms, we
arrive at
\begin{eqnarray}\label{classical_master2}
\dot{\phi}(\sigma,t) &=&\sum_{\sigma'}\left[
P^{-\frac{1}{2}}_{\mathrm{eq}}(\sigma)w(\sigma'\rightarrow \sigma)
P^{\frac{1}{2}}_{\mathrm{eq}}(\sigma')\right.\nonumber\\
&&\left.\hspace{0.8cm}-P^{-\frac{1}{2}}_{\mathrm{eq}}(\sigma')\sum_{\sigma''}w(\sigma'\rightarrow
\sigma'')P^{\frac{1}{2}}_{\mathrm{eq}}(\sigma')\,\delta_{\sigma\sigma'}\right]\phi(\sigma',t)
\nonumber\\
&\equiv&-\sum_{\sigma'}H_{\sigma\sigma' }\phi(\sigma',t)
\end{eqnarray}
with $H$ denoting the real matrix with elements given by
\begin{eqnarray}\label{HamilClassical}
H_{\sigma\sigma'}&=&\sum_{\sigma''}w(\sigma''\rightarrow
\sigma')\,\delta_{\sigma\sigma'}-
P^{-\frac{1}{2}}_{\mathrm{eq}}(\sigma)w(\sigma'\rightarrow
\sigma)P^{\frac{1}{2}}_{\mathrm{eq}}(\sigma').
\end{eqnarray}
Due to the detailed balance condition (\ref{detailed_balance}) one
sees that $H$ is symmetric and hence Eq. (\ref{classical_master2})
gives the aforementioned Schr\"odinger equation.

Diagonalization of the Hamiltonian $H$ is equivalent to the full
solution of the corresponding master equation
(\ref{classical_master}). Many of the previously discussed systems
were investigated from this point of view. Below we recall some of
the known results for the one--dimensional Glauber model. The
Hamiltonian associated to Glauber's master equation
(\ref{classical_master2}) with spin rates given by
(\ref{spinrates}) has the form
\begin{eqnarray}\label{HamilFeld}
H(\gamma,\delta)&=&-\Gamma\sum_{i}\left\{
\left[A(\gamma,\delta)-B(\gamma,\delta)
\sigma_{i-1}^{z}\sigma_{i+1}^{z}\right]\sigma_{i}^{x}\right.\nonumber\\
&&\hspace{1.5cm}\left.-(1+\delta\sigma_{i-1}^{z}\sigma_{i+1}^{z})
[1-\textstyle\frac{\gamma}{2}\sigma_{i}^{z}(\sigma_{i-1}^{z}+\sigma_{i+1}^{z})]\right\},
\end{eqnarray}
where $\gamma$, $\delta$, and $\Gamma$ are specified as
previously; $\sigma^{z}$ and $\sigma^{x}$ are the standard Pauli
matrices and
\begin{equation}\label{AB}
A(\gamma,\delta)=\frac{(1+\delta)\gamma^{2}}{2(1-\sqrt{1-\gamma^{2}})}-\delta,\qquad
B(\gamma,\delta)=1-A(\gamma,\delta).
\end{equation}
For $\delta=0$ it was shown in Ref. \cite{Felderhof} (see also
Ref. \cite{Siggia} for another approach) that the standard
procedure consisting of a Jordan--Wigner transformation followed
by Fourier and Bogoliubov--Valatin transformations
\cite{Bogoliubov,Valatin} brings the Hamiltonian $H(\gamma,0)$ to
its diagonal form with non-interacting fermions. The eigenvalues
are given by
\begin{equation}
\Lambda(\boldsymbol{q}_{k})=\sum_{i=1}^{k}\lambda_{q_{i}}, \qquad \lambda_{q}=1-\gamma\cos q
\end{equation}
where $\boldsymbol{q}_{k}$ denotes the ordered chain $(q_{1}<q_{2}<\ldots<q_{k})$
with each $q_{i}$ from $\{\pm\pi/N,\pm3\pi/N,\ldots,\pm(N-1)\pi/N\}$ for even $N$
and $\{0,\pm2\pi/N,\pm4\pi/N,\ldots,\pi\}$ for odd $N$. The ground state, which is here
the zero--energy state, is the one given by Eq. (\ref{ground}). Moreover, the first nonzero
eigenvalue is $1-\gamma$ and goes to zero for $\gamma\to 1$ (zero temperature limit).
%
%
%

The fermionic representation allows to go beyond Glauber's seminal
work and solve exactly a whole class of master equations that are
associated to the Hamiltonians given by Eq. (\ref{HamilFeld}). The
inverse approach has also been explored: Given a quantum
Hamiltonian that can be solved (at least partially), what are the
corresponding classical master equations, and do they represent
interesting physical systems \cite{droz}?

\section{Quantum kinetic Ising models}
\label{QuantumIsing}

Now we are in a position to proceed with the quantum
generalization of the classical kinetic equations
(\ref{classical_master}). Here we only discuss such a
generalization for the Glauber master equation (\ref{1DGlauber2})
with single--spin flips. However, further generalizations are
obviously possible and are left for future work (see also Sec.
\ref{conclusions}). We start by defining our notation. Naturally,
as the computational basis in $(\mathbb{C}^{2})^{\ot N}$ we take
the eigenstates of $N$-fold tensor product of $\sigma^{z}$. The
basis consists of $2^{N}$ vectors hereafter denoted by
$\ket{\sigma}\equiv \ket{\sigma_{1},\ldots, \sigma_{N}}$
($\sigma=0,\ldots,2^{N}-1$) with $\sigma_{i}$ denoting an
eigenvalue of $\sigma_{i}^{z}$. After appropriate rescaling we may
look at $(\sigma_{1},\ldots, \sigma_{N})$ as the binary
representation of the decimal number $\sigma$.

Let us consider the following master equation
\begin{eqnarray}\label{Qmaster}
\frac{\mathrm{d}\varrho(t)}{\mathrm{d}t}&=&\sum_{i}
\left\{\sigma_{i}^{x}[w_{i}(\sigma^{z})]^{\frac{1}{2}}\,
\varrho(t)[w_{i}(\sigma^{z})]^{\frac{1}{2}}\,\sigma_{i}^{x}
-\frac{1}{2}\left\{w_{i}(\sigma^{z}),\varrho(t)\right\}\right\}.
\end{eqnarray}
Here $\{\cdot,\cdot\}$ denotes the anticommutator and
$w_{i}(\sigma^{z})$ are quantum mechanical generalizations of spin
rates introduced already in the preceding sections\footnote{For
the sakes of clarity and simplicity in the case of quantum spin
rates we change sightly the notation from $w(\sigma\to
D_{i}\sigma)$ to $w_{i}(\sigma^{z})$. According to this convention
$w(D_{i}\sigma\to\sigma)$ is replaced with
$w_{i}(D_{i}\sigma^{z})$.} (the Ising variables $\sigma=\pm 1$ are
replaced by the Pauli matrix $\sigma^{z}$). Thus, all the spin
rates $w_{i}(\sigma)$ $(i=1,\ldots,N)$ are diagonal in the
computational basis $\ket{\sigma}$. It is then clear that the
diagonal part of Eq. (\ref{Qmaster}) reproduce the kinetic
equations (\ref{classical_master}) for all configurations
$\sigma$.
%
%
Notice that Eq. (\ref{Qmaster}) can be written as
\begin{equation}
\frac{\mathrm{d}\varrho(t)}{\mathrm{d}t}=\sum_{i}
\left(L_{i}\varrho(t)L_{i}^{\dagger}-\frac{1}{2}\{L_{i}^{\dagger}L_{i},\varrho(t)\}\right)
\end{equation}
%
with the Lindblad operators given by
$L_{i}=\sigma_{i}^{x}[w_{i}(\sigma^{z})]^{1/2}$. It is then clear
that we deal only with the dissipative part of the general master
equation describing Markov processes.
%

Nevertheless, further generalizations to the full master equation
are possible (see e.g. Refs. \cite{Heims} and \cite{kawasaki}) as,
for instance, for the Ising Hamiltonian, the part governed by the
Hamiltonian vanishes and such an equation would also reproduce the
kinetic equations (\ref{classical_master}).


In what follows we consider only the spin rates that satisfy
detailed balance, i.e.,
\begin{equation}\label{detailed}
w_{i}(\sigma^{z})=
w_{i}(D_{i}\sigma^{z})\exp[-2\beta
J\sigma_{i}^{z}(\sigma_{i-1}^{z}+\sigma_{i+1}^{z})].
\end{equation}
Let us now proceed with solving the above master equation. As we
will see below, our method relies -- as in the case of classical
kinetic equations -- on the observation that the whole equation
can be brought to a set of $2^{N}$ Schr\"odinger equations. For
this purpose it is convenient to use the isomorphism between
matrices from $M_{d}(\mathbb{C})$ and vectors from
$\mathbb{C}^{d}\ot\mathbb{C}^{d}$. More precisely, we can
represent the density matrix $\varrho(t)$ as
\begin{equation}\label{isomorphism}
\varrho(t)=\sum_{\sigma,\widetilde{\sigma}}
[\varrho(t)]_{\sigma,\widetilde{\sigma}}\ket{\sigma}\!\bra{\widetilde{\sigma}}
\quad\longleftrightarrow\quad
\ket{\varrho(t)}=\sum_{\sigma,\widetilde{\sigma}}[\varrho(t)]_{\sigma,\widetilde{\sigma}}\ket{\sigma}
\ket{\widetilde{\sigma}}.
\end{equation}
This form of $\varrho(t)$ allows to rewrite the master equation
(\ref{Qmaster}) as the following matrix equation
\begin{eqnarray}\label{vector_eq}
\ket{\dot{\varrho}(t)}&=&\sum_{i}\left\{\sigma_{i}^{x}
\widetilde{\sigma}_{i}^{x} \sqrt{w_{i}(\sigma^{z})
w_{i}(\widetilde{\sigma}^{z})}-\frac{1}{2}
\left[w_{i}(\sigma^{z})+
w_{i}(\widetilde{\sigma}^{z})\right]\right\}\ket{\varrho(t)},\nonumber\\
\end{eqnarray}
It should be emphasized that operators corresponding to the
"tilded" and "nontilded" spins act on the "tilded" and "nontilded"
kets in the vectors representation of the density matrix
$\varrho(t)$ (\ref{isomorphism}) (for instance
$\sigma_{i}^{x}\widetilde{\sigma}_{i}^{x}\ket{s}\ket{\widetilde{s}}=
\sigma_{i}\ket{s}\widetilde{\sigma}_{i}^{x}\ket{\widetilde{s}}$).

It is evident that the matrix appearing on the right-hand side of
Eq. (\ref{vector_eq}) is not Hermitian. In order to bring it to
Hermitian form we can use the detailed balance condition
(\ref{detailed}). This suggests the following transformation (see
e.g. Ref. \cite{Felderhof})
\begin{equation}\label{transformation}
\ket{\varrho(t)}=
\exp\{-(\beta/4)[\mathcal{H}(\sigma)+
\mathcal{H}(\widetilde{\sigma})]\}\ket{\psi(t)},
\end{equation}
with $\mathcal{H}$ denoting the quantum generalization of the Ising Hamiltonian
given by Eq. (\ref{IsingHamil}). Application of the above to Eq. (\ref{vector_eq})
leads us to
\begin{eqnarray}
\ket{\dot{\psi}(t)}&=&\sum_{i}
\left\{\sigma_{i}^{x}\widetilde{\sigma}_{i}^{x}
[v_{i}(\sigma^{z})]^{\frac{1}{2}}[v_{i}(\widetilde{\sigma}^{z})]^{\frac{1}{2}}
-\frac{1}{2}[w_{i}(\sigma^{z})+w_{i}(\widetilde{\sigma}^{z})]\right\}\ket{\psi(t)},\nonumber\\
\end{eqnarray}
where $v_{i}(\sigma^{z})=w_{i}(\sigma^{z})\exp[\beta
J\sigma_{i}^{z}(\sigma_{i-1}^{z}+\sigma_{i+1}^{z})]$. Now, due to
the detailed balance condition, it is clear that $\sigma_{i}^{x}$
and $v_{i}(\sigma^{z})$ commute, and consequently we have a
Schr\"odinger equation $\ket{\dot{\psi}(t)}=-H\ket{\psi(t)}$ with
Hermitian $H$. Let us notice that the above procedure, being just
a "quantum" generalization of the procedure described in Sec.
\ref{Acciociated}, replaces the problem of solving the master
equation for $N$ spins with the problem of solving the
corresponding Schr\"odinger equation for $2N$ spins.

As we will see below the above equation can be further simplified.
Namely, identifying operators that commute with $H$ we can split
the equation to a group of $2^{N}$ Schr\"odinger equations. For
this aim let us notice that $H$ commutes with
$\sigma_{i}^{z}\widetilde{\sigma}_{i}^{z}$ $(i=1,\ldots,N)$. We
can therefore introduce new variables
$\tau_{i}=\sigma_{i}^{z}\widetilde{\sigma}_{i}^{z}$
$(i=1,\ldots,N)$ which are constants of the motion. Then, one sees
that the tilded variables can be expressed by
$\tau=(\tau_{1},\ldots,\tau_{N})$ and $\sigma$ variables as
$\widetilde{\sigma}_{i}^{z}=\tau_{i}\sigma_{i}^{z}$ for any $i$.
In other words we have replaced $\sigma$ and $\widetilde{\sigma}$
by $\tau$ and $\sigma$, of which $\tau$ is conserved. The
Hamiltonian $H_{\tau}$ then takes  the form
\begin{eqnarray}\label{Htau}
H_{\tau}&=&-\sum_{i}
\left\{\sigma_{i}^{x}[v_{i}(\sigma^{z})]^{\frac{1}{2}}[v_{i}(\tau\sigma^{z})]^{\frac{1}{2}}
-\frac{1}{2}[w_{i}(\sigma^{z})+w_{i}(\tau\sigma^{z})]\right\},
\end{eqnarray}
where $\tau\sigma^{z}$ denotes $\tau_{i}\sigma_{i}^{z}$
$(i=1,\ldots,N)$. Let us make a comment on the notation. One sees
that there are $2^{N}$ different configurations of $\tau$--spins.
%
We label them by the natural numbers $0,\ldots,2^{N}-1$; since we
shall repeatedly have to refer to the two fully homogeneous
configurations we simply denote these as $\tau=0$ (all
$\tau$--spins up) and $\tau=2^{N}-1$ (all $\tau$--spins down),
without further specifying the association of naturals to
configurations.

By definition the $\tau=0$ configuration represents the equal
values of $\sigma$ and $\widetilde{\sigma}$ spins. Consequently,
{\it via} Eq. (\ref{isomorphism}), the corresponding Schr\"odinger
equation describes the diagonal elements of the master equation
(\ref{Qmaster}) and thus recovers the classical kinetic equations
(\ref{1DGlauber2}). In the remaining cases of $\tau$
configurations the $\sigma$ and $\widetilde{\sigma}$ variables
have to differ on at least one position implying that the
Schr\"odinger equation related to any $\tau\neq 0$ describes the
set of $2^{N}$ off--diagonal elements of $\varrho(t)$.
%
%


Thus, by identifying $N$ operators commuting with $H$ we brought
the solution of the general master equation (\ref{Qmaster}) to the
problem of diagonalization of $2^{N}$ Hamiltonians, each of
dimension $2^{N}\times 2^{N}$.

Let us now concentrate on a particular choice for the transition
probabilities $w_{i}(\sigma^{z})$. In what follows we investigate
the quantum version of the rates given by (\ref{spinrates})
($\sigma_{i}$ are replaced with $\sigma_{i}^{z}$), that is
\begin{equation}
w_{i}(\sigma^{z})=\Gamma(1+\delta\sigma_{i-1}^{z}\sigma_{i+1}^{z})
[1-\textstyle\frac{\gamma}{2}\sigma_{i}^{z}(\sigma_{i-1}^{z}+\sigma_{i+1}^{z})]
\end{equation}
with $\Gamma,\delta$, and $\gamma$ defined as before. Putting this
to Eq. (\ref{Htau}) and after some algebra we get
\begin{eqnarray}\label{Htau2}
H_{\tau}(\gamma,\delta)&=&-\Gamma\sum_{i}\left\{
\left[\widetilde{A}_{i}(\gamma,\delta)-\widetilde{B}_{i}(\gamma,\delta)
\sigma_{i-1}^{z}\sigma_{i+1}^{z}\right]\sigma_{i}^{x}-1\right.\nonumber\\
&&\hspace{1.6cm}+\frac{\gamma}{2}(1+\delta)
\sigma_{i}^{z}\left[f(\tau_{i-1}\tau_{i})\sigma_{i-1}^{z}+
f(\tau_{i}\tau_{i+1})\sigma_{i+1}^{z}\right]\nonumber\\
&&\left.\hspace{1.6cm}-\delta f(\tau_{i-1}\tau_{i+1})
\sigma_{i-1}^{z}\sigma_{i+1}^{z}\right\},
\end{eqnarray}
where
\begin{eqnarray}
\widetilde{A}_{i}(\gamma,\delta)=\left\{
\begin{array}{ll}
A(\gamma,\delta), & \tau_{i-1}=\tau_{i+1},\\[3ex]
\sqrt{1-\delta^{2}}\sqrt[4]{1-\gamma^{2}}, &
\tau_{i-1}=-\tau_{i+1}
\end{array}
\right.
\end{eqnarray}
and
\begin{eqnarray}
\widetilde{B}_{i}(\gamma,\delta)=\left\{
\begin{array}{ll}
B(\gamma,\delta), & \tau_{i-1}=\tau_{i+1},\\[3ex]
0, & \tau_{i-1}=-\tau_{i+1}.
\end{array}
\right.
\end{eqnarray}
with $A(\gamma,\delta)$ and $B(\gamma,\delta)$ defined as in Eq.
(\ref{AB}) and $f(x)=(1/2)(1+x)$. One sees that for $\tau=0$ or
$\tau=2^{N}-1$ the function $f$ equals one for $i=1,\ldots,N$ and
therefore we obtain the Hamiltonian given in Eq.
(\ref{HamilFeld}). Let us stress again that the Schr\"odinger
equation corresponding to $\tau=0$ describes the diagonal elements
of the master equation (\ref{Qmaster}) and thus gives the
classical kinetic equations (\ref{1DGlauber2}). All the remaining
$\tau$s correspond to the off--diagonal elements of $\varrho(t)$.

For the interesting special case $\delta=0$ all Hamiltonians
$H_{\tau}$ can be diagonalized using the results of Ref.
\cite{Lieb}. This is because for any $\tau=0,\ldots,2^{N}-1$, the
Jordan--Wigner transformation \cite{JordanWigner} brings
$H_{\tau}(\gamma,\delta)$ to Hamiltonians which are quadratic in
the fermion operators
\begin{equation}
c_{i}=-\mathrm{i}\left(\prod_{j=1}^{i-1}\sigma_{j}^{x}\right)\sigma_{i}^{+},
\qquad \sigma_{i}^{+}=\frac{1}{2}\left(\sigma_{i}^{y}+\mathrm{i}\sigma_{i}^{z}\right).
\end{equation}
This, by virtue of the results of \cite{Lieb}, means that
diagonalization of any of $H_{\tau}(\gamma,0)$ reduces to the
diagonalization of an $N\times N$ matrix. Since, in principle, the
latter can be performed numerically efficiently, the Hamiltonians
(\ref{Htau2}) are diagonalizable.

For the case of $\delta\neq 0$ the Jordan--Wigner transformation
gives us Hamiltonians containing terms which are quartic in
$c_{i}$. In this case we can try the transformation proposed by
Siggia \cite{Siggia}. For this aim we introduce new bond variables
$Z_{i}=\sigma_{i}^{z}\sigma_{i-1}^{z}$,
$\sigma_{i}^{x}=X_{i-1}X_{i}$, and $Y_{i}=-\mathrm{i}Z_{i}X_{i}$.
They commute when their indices are different (different bonds)
whereas for coinciding indices (same bond) they obey the algebra
of Pauli matrices ($X_i^2=1,\, X_iY_i=\mathrm{i}Z_i$ etc). The
bond variables allow to rewrite the Hamiltonian from Eq.
(\ref{Htau2}) as
\begin{eqnarray}
H_{\tau}(\gamma,\delta)&=&-\Gamma\sum_{i}\left\{\widetilde{A}_{i}(\gamma,\delta)X_{i-1}X_{i}
+\widetilde{B}_{i}(\gamma,\delta)Y_{i-1}Y_{i}\right.
\nonumber\\
&&\hspace{1.4cm}-\delta
f(\tau_{i-1}\tau_{i+1})Z_{i-1}Z_{i}\nonumber\\
&&\left.\hspace{1.4cm}-\left[1-\gamma(1+\delta)f(\tau_{i-1}\tau_{i})Z_{i}\right]\right\}.
\end{eqnarray}
Generally this is the anisotropic Heisenberg Hamiltonian with some
external field. In some particular cases such Hamiltonians are
analytically diagonalizable (for a pedagogical review on the
Heisenberg model and the Bethe ansatz that is used to solve it see
Ref. \cite{models}). It seems that the general case $\delta\neq 0$
defies exact diagonalization. However, in the following subsection
we show that for $\delta=-1$, $H_{\tau}(\gamma,\delta)$ can be
diagonalized analytically for all values of $\tau$.

\section{Properties of the Hamiltonians associated to the QME}

Here we study some of the properties of the Hamiltonians
$H_{\tau}$. First, we show that all of them are positive and in a
majority of cases even strictly positive. We also discuss the
cases in which $H_{\tau}$ have zero eigenvalues and thus identify
possible stationary states of the master equation (\ref{Qmaster}).
Then we study fully soluble case of the so--called energy
conserving spin flips, where all Hamiltonians $H_{\tau}$
$\tau=0,\ldots,2^{N}-1$ are diagonalizable. Finally, we
numerically investigate properties of entropy of ground states of
our Hamiltonians.

\subsection{Zero eigenvalues of $H_{\tau}$ --
stationary states of the evolution}

It is interesting to ask which states survive the evolution, that
is, which of the off--diagonal elements of $\varrho(t)$ do not
decay to zero with $t\to\infty$. This can be done by studying the
positivity properties of $H_{\tau}$ and in particular their
eigenstates corresponding to zero eigenvalues. Generically, as we
will see below, for any $\tau$ except for $\tau=0$ and
$\tau=2^{N}-1$, the Hamiltonians $H_{\tau}$ are strictly positive
except for two different $(\gamma,\delta)$--points, namely,
$\delta=\gamma=0$ and $\delta=\gamma=1$. On the other hand, for
$\tau=0$ or $\tau=2^{N}-1$ it is known \cite{Felderhof,HaakeThol}
that the corresponding Hamiltonian has zero eigenvalues (two--fold
degenerate) for any value of $\gamma$ and $\delta$.

Let us treat the two cases  $\tau\neq 0$ and $\tau\neq 2^{N}-1$ in
a more detailed way. For this purpose let us notice that we may
write $H_{\tau}(\gamma,\delta)$ as
$H_{\tau}(\gamma,\delta)=\sum_{i}H_{\tau}^{(i)}(\gamma,\delta)$
and study positivity of each $H_{\tau}^{(i)}(\gamma,\delta)$. On
the other hand, for any $\tau\neq 0,2^{N}-1$ we may divide all
such terms into two groups, the one consisting of
$H_{\tau}^{(i)}(\gamma,\delta)$ for which $\tau_{i-1}=\tau_{i+1}$
and the one for which $\tau_{i-1}=-\tau_{i+1}$. More precisely, we
can write $H_{\tau}(\gamma,\delta)$ as
\begin{equation}
H_{\tau}(\gamma,\delta)=\sum_{\{i|\tau_{i-1}=\tau_{i+1}\}}H_{\tau}^{(i,=)}(\gamma,\delta)
+\sum_{\{i|\tau_{i-1}=-\tau_{i+1}\}}H_{\tau}^{(i,\neq)}(\gamma,\delta).
\end{equation}

In the first case of equal $\tau$s we have
\begin{eqnarray}\label{Hamiltonian_eq}
H_{\tau}^{(i,=)}(\gamma,\delta)&=&\Gamma\left\{1-\textstyle\frac{\gamma}{2}(1+\delta)
f(\tau_{i-1}\tau_{i})\sigma_{i}^{z}\left(\sigma_{i-1}^{z}+
\sigma_{i+1}^{z}\right)+\delta\sigma_{i-1}^{z}\sigma_{i+1}^{z}\right.\nonumber\\
&&\left.\hspace{0.5cm}-\left[A_{i}(\gamma,\delta)
-B_{i}(\gamma,\delta)
\sigma_{i-1}^{z}\sigma_{i+1}^{z}\right]\sigma_{i}^{x}\right\}.
\end{eqnarray}
One finds that $H_{\tau}^{(i)}(\gamma,\delta)$ has four different
eigenvalues, each two--fold degenerate. For $-1\leq \delta\leq 1$
their explicit forms are $\lambda_{1}=0$,
$\lambda_{2}=2\Gamma(1-\delta)$,
\begin{equation}
\lambda_{3,4}^{(=)}=\Gamma(1+\delta)
\left[1\pm\sqrt{1-\gamma+\left(\textstyle\frac{1-\tau_{i-1}\tau_{i}}{2}\right)^{2}\gamma^{2}}\right].
\end{equation}
For the chosen parameter region $\lambda_{2}^{(=)}\geq 0$, while
$\lambda_{3}^{(=)}$ is manifestly positive. On the other hand, to
prove nonnegativity of $\lambda_{4}^{(=)}$ it suffices to notice
that $1-\gamma+[(1-\tau_{i-1}\tau_{i})/2]^{2}\gamma^{2}\leq
1-\gamma+\gamma^{2}\leq 1$ for $\gamma\leq 1$.

In the case of $\tau_{i-1}=-\tau_{i+1}$ we have
\begin{eqnarray}\label{Hamiltonian_neq}
H_{\tau}^{(i,\neq)}(\gamma,\delta)&=&\Gamma\left\{1-\textstyle\frac{\gamma}{2}(1+\delta)\sigma_{i}^{z}
\left(f(\tau_{i-1}\tau_{i})\sigma_{i-1}^{z}+f(\tau_{i}\tau_{i+1})\sigma_{i+1}^{z}\right)
\right.\nonumber\\
&&\left.\hspace{0.5cm}-\sqrt{1-\delta^{2}}\,\sqrt[4]{1-\gamma^{2}}\,\sigma_{i}^{x}\right\}.
\end{eqnarray}
Here one finds that since $\tau_{i-1}=-\tau_{i+1}$, one of the
factors $f(\tau_{i-1}\tau_{i})$ or $f(\tau_{i}\tau_{i+1})$
vanishes. Without any loss of generality let us assume that
$f(\tau_{i}\tau_{i+1})=0$. Then, obviously
$f(\tau_{i-1}\tau_{i})=1$ and the above Hamiltonian can be brought
to
\begin{eqnarray}
H_{\tau}^{(i,\neq)}(\gamma,\delta)&=&1-\textstyle\frac{\gamma}{2}(1+\delta)\sigma_{i-1}^{z}\sigma_{i}^{z}
-\sqrt{1-\delta^{2}}\,\sqrt[4]{1-\gamma^{2}}\,\sigma_{i}^{x}.
\end{eqnarray}
It has two different eigenvalues (each two--fold degenerate) of
the form
\begin{equation}\label{eigenvalues_neq}
\lambda_{\pm}^{(\neq
)}=1\pm\frac{1}{2}\sqrt{\gamma^{2}(1+\delta)^{2}+4\sqrt{1-\gamma^{2}}(1-\delta^{2})}.
\end{equation}
One sees that obviously $\lambda_{+}^{(\neq )}\geq 0$. To see that
also $\lambda_{-}^{(\neq )}$ is nonnegative it suffices to notice
that for $0\leq \gamma\leq 1$ and $|\delta|\leq 1$ the maximal
value of the function of $\gamma$ and $\delta$ appearing under the
sign of square root is four and is this value is attained for
$\gamma=\delta=0$ and $\delta=\gamma=1$. This also means that
$\gamma=\delta=0$ and $\gamma=\delta=1$ are the only points for
which $H_{\tau}^{(i,\neq)}(\gamma,\delta)$ can have zero
eigenvalues.

In conclusion, it follows from the above analysis that all
$H_{\tau}^{(i)}(\gamma,\delta)$s are positive and therefore the
Hamiltonian $H_{\tau}(\gamma,\delta)$ is positive for any
$\tau=0,\ldots,2^{N}-1$, and parameter region specified by the
conditions $-1\leq\delta\leq 1$, $0\leq \gamma\leq 1$, and
$\Gamma>0$. Moreover, it follows that for all $\tau\neq
0,2^{N}-1$, the Hamiltonians $H_{\tau}(\gamma,\delta)$ are in
general strictly positive except the points $\delta=\gamma=0$ and
$\delta=\gamma=1$. Thus these two points together with two values
of $\tau=0,2^{N}-1$ are the only possible cases when
$H_{\tau}(\gamma,\delta)$ can have zero eigenvalues. Let us
discuss shortly each of these cases.

For $\tau=0$ and $\tau=2^{N}-1$, as previously noticed, the
corresponding Hamiltonian (\ref{HamilFeld}) has a zero--eigenvalue
eigenstate given by Eq. (\ref{ground}). The case $\tau=0$
($\tau=2^{N}-1$) corresponds to the diagonal (anti--diagonal)
elements of $\varrho(t)$, which as it follows from Eqs.
(\ref{ground}) and (\ref{transformation}) are of the form (taking
into account the normalization)
\begin{equation}
\ket{\varrho_{\tau}(t)}=\frac{1}{Z_{N}}\,\mathrm{e}^{-\beta\mathcal{H}(\sigma)}
\sum_{\sigma}\ket{\sigma}\qquad (\tau=0,2^{N}-1).
\end{equation}
For the zero temperature limit ($\beta\to\infty$), the above
becomes the well--known $N$--partite Schr\"odinger cat state (or
Greenberger--Horne--Zeilinger state)
\begin{equation}
\ket{\psi^{(N)}_{+}}=\frac{1}{\sqrt{2}}\left(\ket{\!\!\uparrow}^{\ot
N}+\ket{\!\!\downarrow}^{\ot N}\right)
\end{equation}
with $\ket{\!\!\uparrow}$ and $\ket{\!\!\downarrow}$ denoting the
eigenvectors of $\sigma^{z}$ corresponding to the positive and
negative eigenvalue, respectively. Moreover, for $\gamma=1$ the
two--fold degeneracy appears in the ground state of
$H_{1}(1,\delta)$ and the second zero--energy eigenstate is
\begin{equation}
\ket{\psi^{(N)}_{-}}=\frac{1}{\sqrt{2}}\left(\ket{\!\!\uparrow}^{\ot
N}-\ket{\!\!\downarrow}^{\ot N}\right).
\end{equation}
Since generically the remaining off--diagonal elements of
$\varrho(t)$ vanish in the $t\to\infty$ limit (except for the
already mentioned values of $\gamma$ and $\delta$), we have an
example of a state that is a stationary state of the master
equation and becomes genuine multipartite entangled state in the
limit of zero temperature.

Let us now treat the cases of $\gamma=\delta=0$ and
$\delta=\gamma=1$. In the first one, the Hamiltonians
$H_{\tau}(\gamma,\delta)$ simplify significantly and are of the
form
\begin{equation}
H_{\tau}(0,0)=\Gamma\sum_{i}(1-\sigma_{i}^{x})\qquad
(\tau=0,\ldots,2^{N}-1),
\end{equation}
meaning that the only zero energy state is $\ket{\!\!\to}^{\ot N}$
where $\ket{\!\!\rightarrow}$ and $\ket{\!\!\leftarrow}$ stand for
the normalized eigenstates of $\sigma^{x}$ corresponding to
positive and negative eigenvalue, respectively. Since, as it
follows from Eq. (\ref{transformation}), for $\gamma=1$ it holds
that $\ket{\varrho_{\tau}(t)}=\ket{\psi_{t}(t)}$ for any $\tau$
and therefore
$\ket{\varrho_{\tau}(t)}=(1/\sqrt{2^{N}})\ket{\!\to}^{\ot N}$ for
$\tau=0,\ldots,2^{N}-1$. Consequently, the stationary state is
fully separable and is given by
$\varrho_{\mathrm{st}}(t)=P_{\ket{\to}}^{\ot N}$ with
$P_{\ket{\to}}$ denoting a projector onto $\ket{\!\!\to}$.

The case of $\delta=\gamma=1$ is a little bit more difficult. Now,
from Eq. (\ref{Htau2}) it follows that
\begin{eqnarray}\label{Htau3}
H_{\tau}(1,1)&=&\Gamma\sum_{i}\left\{1-
\sigma_{i}^{z}\left[f(\tau_{i-1}\tau_{i})\sigma_{i-1}^{z}+
f(\tau_{i}\tau_{i+1})\sigma_{i+1}^{z}\right]\right.\nonumber\\
&&\left.\left.\hspace{1.0cm}+f(\tau_{i-1}\tau_{i+1})
\sigma_{i-1}^{z}\sigma_{i+1}^{z}\right]\right\}.
\end{eqnarray}
It is clear that this Hamiltonian is diagonal in the standard
basis $\ket{\sigma}$ in $(\mathbb{C}^{2})^{\ot N}$ and thus we can
look for the eigenstates among the standard basis in
$(\mathbb{C}^{2})^{\ot N}$. Interestingly, using the previously
introduced bond variables $Z_{i}=\sigma_{i-1}^{z}\sigma_{i}^{z}$,
in the case of periodic boundary conditions this Hamiltonian can
be brought to the antiferromagnetic Ising Hamiltonian with
magnetic field
\begin{equation}
H_{\tau}(1,1)=\Gamma
N+\sum_{i}f(\tau_{i-1}\tau_{i+1})Z_{i-1}Z_{i}-2\Gamma\sum_{i}f(\tau_{i-1}\tau_{i})Z_{i}.
\end{equation}

Let us concentrate on the zero--energy eigenstates of
$H_{\tau}(1,1)$. The latter can be found by solving the
corresponding equation for eigenvalues of $\sigma^{z}$. This,
however, due to the fact that in this equation each term under the
sum is positive, means solving of the following set of equations
\begin{eqnarray}
&&f(\tau_{i-1}\tau_{i})\sigma_{i-1}\sigma_{i}+f(\tau_{i}\tau_{i+1})\sigma_{i}\sigma_{i+1}-
f(\tau_{i-1}\tau_{i+1})\sigma_{i-1}\sigma_{i+1}=1\nonumber\\
&&\hspace{7.5cm}(i=1,\ldots,N),
\end{eqnarray}
where $\sigma_{i}$ stands for the eigenvalue of $\sigma_{i}^{z}$.
For instance, for all configurations of $\tau$ that consist of
blocks of length greater or equal two separated by the domain
walls, one of the possible solutions is given by
$\sigma_{i}=\tau_{i}$. This is because in such case the above set
becomes
$f(\tau_{i-1}\tau_{i})+f(\tau_{i}\tau_{i+1})-f(\tau_{i-1}\tau_{i+1})=1$
$(i=1,\ldots,N)$. The only possible triples
$(\tau_{i-1},\tau_{i},\tau_{i+1})$ that can appear in the
discussed case are $\uparrow\uparrow\uparrow$,
$\downarrow\downarrow\downarrow$, $\uparrow\downarrow\downarrow$,
$\downarrow\uparrow\uparrow$, $\uparrow\uparrow\downarrow$, and
$\downarrow\downarrow\uparrow$. It is clear that for all of them
these equations are satisfied. One may also easily check that
generically the zero--energy eigenstates are degenerated. As a
result it seems that in the case of $\gamma=\delta=1$ there is a
variety of states $\varrho(t)$ that are stationary states of our
master equation.

\subsection{The case of energy conserving spin flips}

Here we consider the case of energy--conserving spin flips, that is, when
$\delta=-1$. One sees now that the Hamiltonians $H_{\tau}$ simplify
significantly and read
\begin{eqnarray}\label{Heisenberg-1}
H_{\tau}(\gamma,-1)=\Gamma
N-\Gamma\sum_{i}f_{i}^{\tau}\left(X_{i-1}X_{i}
+Y_{i-1}Y_{i}+Z_{i-1}Z_{i}\right),
\end{eqnarray}
where $f_{i}^{\tau}=f(\tau_{i-1}\tau_{i+1})$. For $\tau=0$ and
$\tau=2^{N}-1$ this is the isotropic ferromagnetic Heisenberg
model with spectrum shifted by $\Gamma N$. In the general case,
however, $H_{\tau}(\gamma,-1)$ depends on the numbers $f_{i}$
which are either zero or one depending on the configuration
$\tau$. To deal with this it suffices to notice that for any
configuration of $\tau\neq 0,2^{N}-1$  some set of indices
$\{i_{1},\ldots,i_{k}\}$ exists for which $f_{i_{k}}=0$ and
between these zeros the $f_{i}$ are constant and equal to one. It
is clear from Eq. (\ref{Heisenberg-1}) that such zeros divide the
Hamiltonian into a sum of "smaller" commuting Hamiltonians. More
precisely if $f_{i_{j}}=0$ for $j=1,\ldots,k$ then
\begin{eqnarray}
H_{\tau}(\gamma,-1)&=&\Gamma
N-\Gamma\sum_{n=1}^{i_{1}-1}\boldsymbol{S}_{n-1}\cdot
\boldsymbol{S}_{n}-\Gamma\sum_{n=i_{1}+1}^{i_{2}-1}\boldsymbol{S}_{n-1}\cdot
\boldsymbol{S}_{n}-\ldots\nonumber\\
&&-\sum_{n=i_{k-1}+1}^{i_{k}-1}\boldsymbol{S}_{n-1}\cdot
\boldsymbol{S}_{n},
\end{eqnarray}
where $\boldsymbol{S}_{n}=[X_{n},Y_{n},Z_{n}]$. It follows that
$\boldsymbol{S}_{n-1}\cdot \boldsymbol{S}_{n}$ commutes with
$\boldsymbol{S}_{m-1}\cdot \boldsymbol{S}_{m}$ whenever
$|n-m|\geq2$.

This means that for a given configuration of $\tau$ we can split
$H_{\tau}(\gamma,-1)$ into a group of commuting isotropic
ferromagnetic Heisenberg Hamiltonians with free ends. Such
Hamiltonians can be treated using the so--called Bethe ansatz
\cite{Bethe}. To visualize what we have just said let us consider
the following illustrative example. Let us assume the periodic
boundary conditions in Eq. (\ref{Heisenberg-1}) and let the $\tau$
configuration together with the corresponding chain $f_{i}^{\tau}$
be given by
\begin{eqnarray}
\begin{array}{c|cccccccccccccc}
\tau & \uparrow & \uparrow & \uparrow & \downarrow & \downarrow &
\uparrow & \uparrow & \uparrow & \downarrow & \uparrow &
\downarrow & \uparrow & \uparrow & \uparrow \\\hline f_{i} & 1 & 1
& 0 & 0 & 0 & 0 & 1 & 0 & 1 & 1 & 1 & 0 & 1 & 1
\end{array}
\end{eqnarray}
Then the Hamiltonian becomes (forgetting about the constant part
and putting $\Gamma=1$)
$H_{\tau}(\gamma,-1)=-(\sum_{i=1}^{2}\boldsymbol{S}_{i-1}\cdot
\boldsymbol{S}_{i}+\boldsymbol{S}_{6}\cdot
\boldsymbol{S}_{7}+\sum_{i=9}^{11}\boldsymbol{S}_{i-1}\cdot
\boldsymbol{S}_{i}+\sum_{i=13}^{14}\boldsymbol{S}_{i-1}\cdot
\boldsymbol{S}_{i})$.

\subsection{Entropy of the ground state of the off--diagonal Hamiltonians}
Here we take a brief detour and study the entanglement between
parts of the ground state of the Hamiltonians
$H_\tau(\gamma,\delta)$ that control the dynamics of the
off--diagonal terms of the QME. In order to study our system for
arbitrary values of $\delta$, we take advantage of the matrix
product state structure that the ground state of these
Hamiltonians has. For this, we extended the time evolving block
decimation (TEBD) algorithm of Vidal \cite{Vidal} to include
next-nearest neighbor interaction. We then performed a variable
step imaginary time evolution to find the ground state of open
boundary chains of length $N$ large enough so that the results
become independent of size. Our goal is to compute the bipartite
entropy $S=\tr \rho_L \log_2 \rho_L$, where $\rho_L$ is the
reduced density matrix obtained after tracing out $N-L$
neighboring spins of the chain. For this, instead of obtaining and
diagonalizing $\rho_L$, we make use of the Schmidt coefficients
that appear explicitly in the TEBD representation of the
state\footnote{ Any pure state $\ket{\psi}$ of a quantum system
partitioned into two parts $A$ and $B$ can be written in its
Schmidt decomposition form, $\ket{\psi}=\sum^\chi_{\alpha=1}
\lambda_\alpha^{1/2}
\ket{\phi_\alpha^{(A)}}\ket{\varphi^{(B)}_\alpha}$, where
$\{\ket{\phi_\alpha^{(A)}}\}$ and $\{\ket{\varphi_\alpha^{(B)}}\}$
are two orthonormal bases of the Hilbert space of parts $A$ and
$B$ respectively, $\lambda_\alpha$ are non-negative real numbers
(the Schmidt coefficients), and $\chi$ is the smallest of the
dimensions of the Hilbert spaces of $A$ and $B$. If part $B$ is
then traced out, the reduced density matrix of part $A$ can be
written as a diagonal matrix in the $\{\ket{\phi_\alpha^{(A)}}\}$
basis, with the Schmidt coefficients in the diagonal. Since the
TEBD algorithm basically stores the Schmidt coefficients and the
Schmidt bases for all possible bi-partitions of the system, then
computing the von-Neumann entropy of the reduced density matrix
$\rho_L$ for any $L$-value is quite easy. }.

\begin{figure}[htbp]
   \centering
   \includegraphics[width=4in]{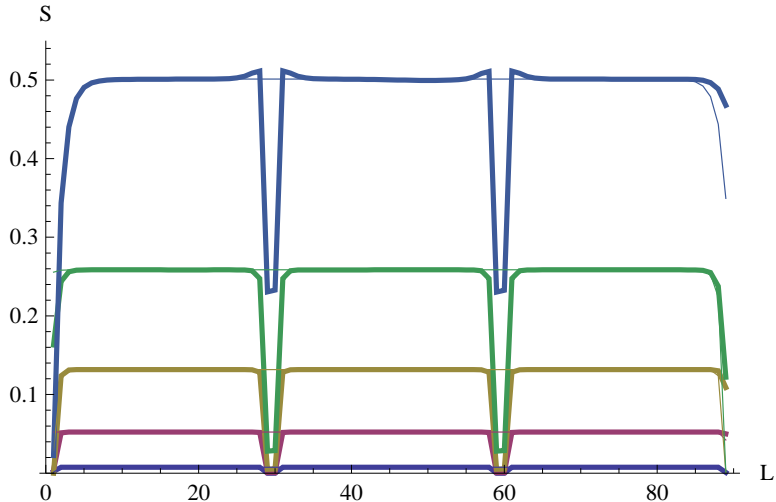}
   \caption{Bipartite entropy $S(L)=\tr \rho_L \log_2 \rho_L$, where $\rho_L$ is the
reduced density matrix obtained after tracing out $N-L$ contiguous
spins of the chain, for the Hamiltonian (\ref{Htau2}) in the
$\delta=0$ case, and where $\tau$ is a configuration where all but
two components are equal. The curves, in ascending order, are for
$\gamma=0.1, 0.3, 0.5, 0.7$, and $0.9$. The entropy reaches $1$
for criticality at $\gamma=1$. The system is a chain of 90 spins.
Notice how at the point of the flipped $\tau_i$'s the bipartite
entropy is reduced but does not go to zero. In thin lines, for
comparison, is the same entropy but for a configuration with all
$\tau_i$'s equal to one (the classical KIM, here the diagonal part
of the QME). }
   \label{Fig:1}
\end{figure}

We study two representative cases of $\tau$ configurations, one in
which only two $\tau_i$'s are different than the rest, and one
where half neighboring $\tau_i$'s are equal between them and
different than the other half - a case that we call a "domain
wall". We also concentrate on two relevant values of $\delta$,
first $\delta=0$ (the Glauber model), shown in Figs. (1) and (2),
and second, the temperature dependent $\delta=\gamma/(2-\gamma)$
(the Haake-Thol model that gives the dynamical critical exponent
$z=4$), shown in Figs. (3) and (4). For comparison, we also
compute the bipartite entropy of the diagonal component of the
QME, i.e., the classical KIM.

\begin{figure}[htbp]
   \centering
   \includegraphics[width=4in]{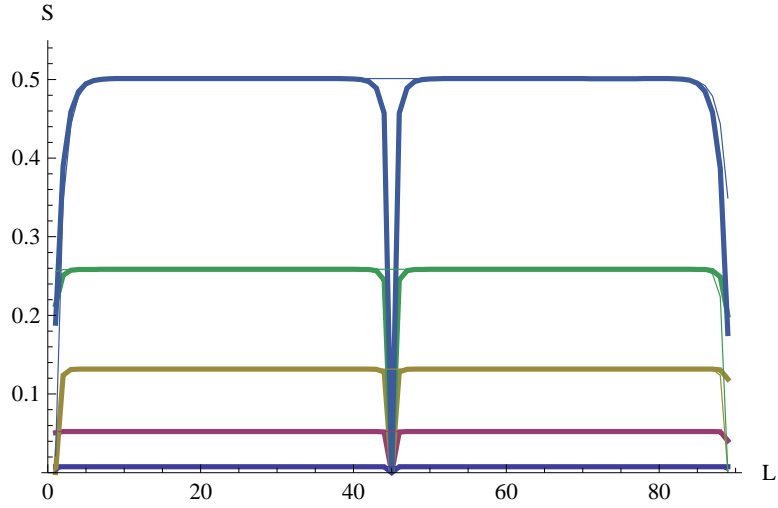}
   \caption{The same as Fig. 1 but for a configuration of $\tau_i$'s
   where half are $1$ and the other half are $-1$. Notice how the entropy
   does go to zero at the domain wall.
}
   \label{Fig:2}
\end{figure}

We observe that the maximum value of entropy grows with $\gamma$
and approaches unity for criticality ($\gamma=1$). The block
length $L$ at which the entropy saturates also appears to be
rather small, about 5 sites, although we expect this to grow near
the critical point.

\begin{figure}[htbp]
   \centering
   \includegraphics[width=4in]{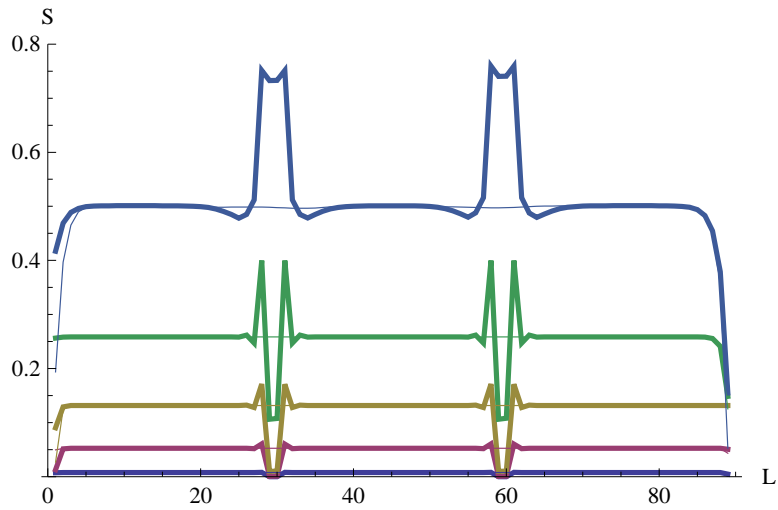}
   \caption{The same system as Fig. 1, but for $\delta=\gamma/(2-\gamma)$,
   where the classical KIM shows an anomalous dynamical exponent $z=4$.
In this case, the flipped $\tau_i$ induces a spike in entropy near the domain wall, indicating
some interesting quantum correlation existing between domain walls.
Still the curves, in ascending order, are for $\gamma=0.1, 0.3, 0.5, 0.7$, and $0.9$.
In thin lines, for comparison, is the same system but for a configuration with all
$\tau_i$'s equal to one (the classical KIM, here the diagonal part of the QME).
}
   \label{Fig:3}
\end{figure}

In all cases of $\delta$ a domain wall appears to de-entangle the
two parts of the chain, giving zero entropy for a partition right
at the domain boundary. On the other hand, a single flipped
$\tau_i$ shows some residual entropy. Therefore, the separation of
the Hamiltonian into commuting parts shown above for $\delta=-1$
is not possible for general $\tau$ configurations.

\begin{figure}[htbp]
   \centering
   \includegraphics[width=4in]{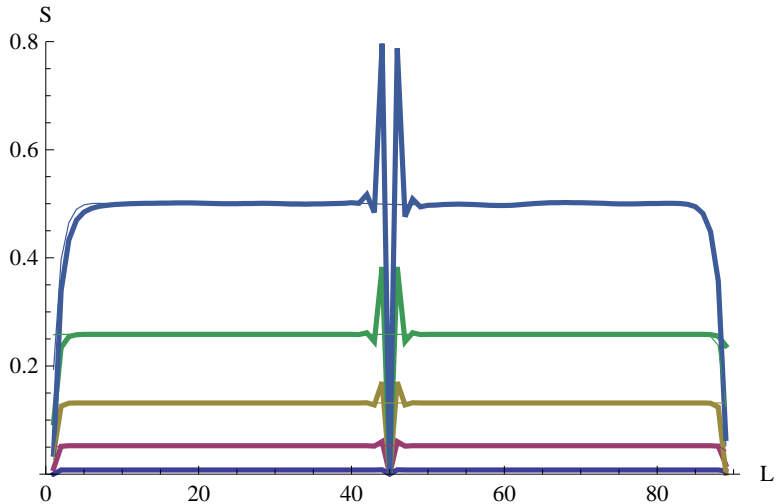}
   \caption{The same system as Fig. 3, but for a domain wall configuration for $\tau$.
   In contrast to the case shown in Fig. 2, here the entropy grows sharply
   before the end of the domain. Nevertheless, it goes to zero exactly at the domain
   wall, again indicating that in this case the domains are not entangled.
}
   \label{Fig:4}
\end{figure}

The case with an anomalous dynamical exponent $z=4$,  $\delta=\gamma/(2-\gamma)$,
shows some interesting behavior of the entropy near the domain walls or the
flipped $\tau_i$. In particular, we observe an entropy spike before the end of the domain.
The effect is acutely pronounced near criticality.

%
\section{Conclusions and Outlook}
\label{conclusions}

We have presented here a novel class of QMEs that have the
following properties (cf. e.g. Refs. \cite{Heims} and
\cite{kawasaki}):
\begin{itemize}
\item The diagonal elements of the density matrix in the "computational" basis
follow  dynamics of a certain classical kinetic model.

\item The dynamics of the off-diagonal matrix elements splits
into blocks, described by "kinetic--like" models.

\item For models fulfilling DBC, the
dynamics can be transformed into a Hamiltonian dynamics in
imaginary time.

\item The ground and low excited states of the resulting
Hamiltonians fulfill area law, and can be well described by MPS or PEPS methods.
\end{itemize}

Our results suggest several directions of investigations, which we
would like to follow in the future:
\begin{itemize}

\item In the present paper we have focused on Ising spins,
and on generalized QME associated with KIMs. Generalizations to
models associated with kinetics of more general set of commuting
operators, such as stabilizer operators, are possible and
interesting.

\item Several presented models admit exact solutions via Wigner-Jordan
transformation or Bethe ansatz \'a la \cite{Siggia,Felderhof},
and/or approximate treatment  using variational methods \'a la
\cite{HaakeThol}. These methods should allow for more rigorous
analysis of the novel types of Hamiltonians.

\item Especially interesting are the two--spin--flip models such as
those that conserve the magnetization \cite{Suzuki} or energy
\cite{Felderhof2}. In particular, for the energy conserving model
in 1D the generalized QME reads:
\begin{eqnarray}\label{Qmaster-ener}
\frac{\mathrm{d}\varrho(t)}{\mathrm{d}t}&=&\sum_{i=1}^{N}
\left[\sigma_{i}^{x}\sigma_{i+1}^{x} \sqrt{1-
\sigma_{i-1}^{z}\sigma_{i}^{z}\sigma_{i+1}^{z}\sigma_{i+2}^{z}}
\,\varrho(t)\right.\nonumber\\
&&\hspace{0.8cm}\times \sqrt{1-
\sigma_{i-1}^{z}\sigma_{i}^{z}\sigma_{i+1}^{z}\sigma_{i+2}^{z}}\,
\sigma_{i}^{x}\sigma_{i+1}^{x}\nonumber\\
&&\left.\hspace{0.8cm}-\frac{1}{2}\left\{1-
\sigma_{i-1}^{z}\sigma_{i}^{z}\sigma_{i+1}^{z}\sigma_{i+2}^{z},\varrho(t)\right\}\right],
\end{eqnarray}
The high degeneracy of the ground state in this model allows to
expect the appearance of long--living coherences.

\item Generalization to models that do not fulfill DBC such as
exclusion models (see Refs. \cite{Darrida,Temme}) is possible. An
example of somewhat "hidden" DBC is the QME of the form
\begin{eqnarray}\label{Qmaster-heis}
\frac{\mathrm{d}\varrho(t)}{\mathrm{d}t}&=&\sum_{i=1}^{N}
\left[\sigma_{i}^{x}\sigma_{i+1}^{x} \sqrt{1-
\alpha\sigma_{i}^{z}\sigma_{i+1}^{z}} \,\varrho(t) \sqrt{1-
\alpha\sigma_{i}^{z}\sigma_{i+1}^{z}}\,
\sigma_{i}^{x}\sigma_{i+1}^{x}\right.\nonumber\\
&&\left.\hspace{0.8cm}-\frac{1}{2}\left\{1-
\alpha\sigma_{i}^{z}\sigma_{i+1}^{z},\varrho(t)\right\}\right],
\end{eqnarray}
where $-1\le \alpha\le 1$ is a free parameter. This model
corresponds to 1D anisotropic Heisenberg model.

\item Last, but not least, physical implementations of the
considered models with ultracold atoms in  optical lattices, or
ions in trap arrays, or Rydberg atoms are feasible, and will be
studied.

\end{itemize}

\section{Acknowledgments}

We thank W. D\"ur, M.-C. Banyuls, and H. Briegel for discussions.
R. A. thanks J. Stasi\'nska for discussions. We acknowledge the
support of Spanish MEC/MINCIN projects TOQATA (FIS2008-00784) and
QOIT (Consolider Ingenio 2010), ESF/MEC project FERMIX
(FIS2007-29996-E), EU Integrated Project  SCALA, EU STREP project
NAMEQUAM, ERC Advanced Grant QUAGATUA, Caixa Manresa, and
Alexander von Humboldt Foundation Senior Research Prize.

\end{document}